\documentclass[
 aip,
 amsmath,
 amssymb,
 reprint,
 jcp,
]{revtex4-1}

\usepackage[utf8]{inputenc}
\usepackage[T1]{fontenc}
\usepackage{mathptmx}
\usepackage{etoolbox}

\usepackage{graphicx}
\usepackage{dcolumn}
\usepackage{bm}
\usepackage{hyperref}

\date{\today}

\begin{document}

\title{Generating candidates in global optimization algorithms using complementary energy landscapes}

\author{Andreas Møller Slavensky}
\author{Mads-Peter Verner Christiansen}
\author{Bjørk Hammer}
    \email{hammer@phys.au.dk}
\affiliation{Center for Interstellar Catalysis, Department of Physics and Astronomy, Aarhus University, DK‐8000 Aarhus C, Denmark}

\begin{abstract}
  Global optimization of atomistic structure rely on the generation of
  new candidate structures in order to drive the exploration of the
  potential energy surface (PES) in search for the global minimum
  energy (GM) structure.  In this work, we discuss a type of structure
  generation, which locally optimizes structures in complementary
  energy (CE) landscapes. These landscapes are formulated temporarily
  during the searches as machine learned potentials (MLPs) using local
  atomistic environments sampled from collected data. The CE
  landscapes are deliberately incomplete MLPs that rather than
  mimicking every aspect of the true PES are sought to become much
  smoother, having only few local minima.  This means that local
  optimization in the CE landscapes may facilitate identification of
  new funnels in the true PES. We discuss how to construct the CE
  landscapes and we test their influence on global optimization of a
  reduced rutile SnO$_2$(110)-(4$\times$1) surface, and an olivine (Mg$_2$SiO$_4$)$_4$ cluster for
  which we report a new global minimum energy structure.
\end{abstract}

\maketitle

\section{Introduction}
A wide range of global optimization methods exist for the
identification of the global minimum energy structures of atomstic
systems, such as molecular aggregates, clusters, and crystals. Some
examples of popular methods are random structure search \cite{RSS2011}
(RSS), basin-hopping \cite{BH1997} (BH), simulated annealing
\cite{simulated_annealing}, minima-hopping
\cite{goedecker_minima_2004} (MH),
particle swarm algorithms \cite{PS2010}, and evolutionary algorithms
\cite{EA2003,deaven_molecular_1995,wu_adaptive_2014} (EA).
When used in conjunction with first-principles methods such as density
functional theory (DFT), the global optimization (GO) methods have proven
efficient in identifying highly elaborate global minimum energy
structures, such as e.g.\ 
crystal structures\cite{oganov_crystal_2006,banerjee_crystal_2021,curtis_gator_2018},
gas-phase clusters\cite{davis_birmingham_2015,paleico_flexible_2020},
surface-supported clusters\cite{vilhelmsen_genetic_2014,jager_giga_2019,penschke_vanadium_2018},
surface defects\cite{stausholm-moller_density_2013}, and surface reconstructions\cite{zakaryan_stable_2017,merte_structure_2017}.

The impressive advances within the field of constructing
machine-learned potentials (MLPs) based on databases of structures
with DFT energies have recently made their way into global
optimization. Most straighforward, the MLPs are constructed on-the-fly
during the GO searches, and structural candidates are propagated, with
molecular dynamics or local relaxation (quenching) on these MLPs. This
approach has e.g.\ lead to the identification of the optimal structure
of different clusters\cite{jennings_genetic_2019,ouyang_global_2015,tong_accelerating_2018,kaappa_global_2021},
crystal structures\cite{podryabinkin_accelerating_2019}, surface adsorbates\cite{liu_combining_2023,todorovic_bayesian_2019,kolsbjerg_neural-network-enhanced_2018,jung_machine-learning_2022}, defects in solids\cite{arrigoni_evolutionary_2021}, and to the unraveling of an intricate
$(4\times 4)$-reconstruction of SnO$_2$ on Pt$_3$Sn(111) \cite{merte_2022}. We
note that such global optimization accelerated by on-the-fly learnt
MLPs aims at identifying the GM structures of matter, while the
concurrent formulation of accurate MLPs can be considered a
biproduct. Contrary to this, the disciplines of active and batch learning of MLPs
target the formulation of accurate MLPs directly. These MLPs may
eventually be used in simulations of various kinds, including
vibrational analysis\cite{gastegger_machine_2017,beckmann_infrared_2022,zhang_modeling_2021}, phase diagram mapping\cite{timmermann_data-efficient_2021,rosenbrock_machine-learned_2021,jinnouchi_phase_2019,erhard_machine-learned_2022}, and of
course also global optimization\cite{chiriki_neural_2017,deringer_data-driven_2018,paleico_global_2020}.

In global optimization, only the end result of actually identifying
the GM structure matters and not the details on how it was found. Thus
more hypothetical models may be formulated as for instance some in which
the dimensionality of the problems is altered with the aim of allowing for more
efficient exploration.
An example of this is the addition of extra spatial dimensions for the
atoms, which allows the local optimization of a structure to escape
energy minima \cite{Hyperspatial2019}. Other examples are the addition of extra atoms
and their generalization into either \textit{fractional} atoms \cite{larsen_beacon_2022} or \textit{mixed} atoms
\cite{kaappa_beacon_2021}.

In our group, we have previously proposed the concept of assisting the
global optimization with formulation of complementary energy (CE)
landscapes and local relaxation of candidate structures in these
landscapes. This approach is useful for GO methods such as RSS, BH,
MH, and EA, that all have in common that they rely on
stochastic candidate production and subsequent local relaxation
(quenching) in the PES. In Ref \onlinecite{Clustering2018}, the \textit{cluster-regularization}
method was presented. Here a CE landscape is constructed by
clustering in a feature space the local atomic environments of some
reference structures. The complementary energy of a given structure
is then obtained as the sum of the shortest distances to the cluster centers
of each of the local atomic environments within the structure. Optimizing
in this CE, the cluster centers effectively act as attractors to the
local environments of the atoms. In Ref \onlinecite{Rolemodel2019}, we formulated the
\textit{role-model} method in which the clustering step is omitted and where attractors are
chosen among actually existing atoms. Recently, the Goedecker group has
proposed a related method in which the PES is modified such that symmetric structures
with few distinct local atomic environments are favored \cite{Symmetry2022}. 

In the present work, we revisit the role-model method in order to
provide a critical discussion of how the CE landscape is best
constructed. Three major choices made for a CE are considered: i) The
descriptor for the local atomic environments, ii) the attractors
(i.e.\ the desired local environments), and iii) the 
expression that converts actual local environments and attractors into
a complementary energy value for a given structure. We find that a
few-dimensional, density-based representation of the atomic
environments is sufficient, that attractors may be based on the actual
structure being stochastically modified (in either Basin Hopping or
Evolutionary Algorithm optimization), and that a complementary energy
expression based on the Euclidian distance between environments and
attractors may be used. These are surprizingly simple choices compared to
the common elements of MLPs, where much more complicated features and
energy expressions are used. This highlights the different role that
CE landscapes are playing compared to MLPs.

The paper is outlined as follows: First, the Complementary Energy method is
described and the effects of different choices of energy expressions,
local descriptors, and attractors are investigated. This is
done for a simple 2-dimensional Sn$_3$O$_6$ cluster, where the CE
landscape and the local environments and attractors can easily be visualized.
Second, the use of a CE landscape in a search context, where the
stochastic perturbation of a candidate is done via relaxation in the
CE landscape, is introduced and illustrated for a test system, a 2-dimensional silicate cluster, (MgSiO$_3$)$_2$.
Third, a machine learning enhanced Basin Hopping and the Global Optimization with First-principles
Energy Expressions\cite{GOFEE2020, GOFEE2022} (GOFEE) search methods are introduced. Both methods rely on Gaussian Process Regression\cite{GPR2005} (GPR)
to construct a MLP, also referred to as a surrogate model. 
Fourth, we consider the globally
optimal surface reconstruction of reduced SnO$_2$(110)-(4$\times$1)
and compare in a Basin Hopping setting the speedup
obtained when introducing the CE method for creating stochastic
perturbations of the candidate structure. Finally, we confirm for a
larger, more complicated problem, an olivine (Mg$_2$SiO$_4$)$_4$
cluster, that the CE method indeed speeds up GO search. For
this system we further identify a new, hitherto not reported geometry
for the GM structure.

The present CE method has been implemented as a structure
generator for use with the Atomistic Global Optimization X
\cite{AGOX2022} (AGOX) Python package, which can be obtained at \href{https://gitlab.com/agox/agox}{https://gitlab.com/agox/agox}.

\section{Complementary energy generator}
\label{sec:complementary_section}

In this work, the complementary energy $CE$ of a structure consisting of $N$ atoms
is a function of the local environments of the atoms and some attractors,
\begin{align}
    CE= h(\textbf{f}_1, \textbf{f}_2, \dots, \textbf{f}_N,\textbf{A}_1,\dots,\textbf{A}_M),
\end{align}
where $\textbf{f}_i$ is the description of the local environment of atom $i$, which is encoded in a feature vector,
and \textbf{A}$_j$ is the feature vector of the $j$'th attractor.
Note that, analogous to their use in highly accurate machine learning
potentials\cite{rupp_fast_2012,miksch_strategies_2021}, descriptors that encode the symmetries of the Hamiltonian can be used to construct the complementary energy landscape.
This ensures that the complementary energy respects the same invariances as the target energy function. 

There exist several ways of how to evaluate the local descriptors,
how to calculate the complementary energy, and how to choose the attractors. In the following sections, some of these possible choices will be discussed, and 
their effect on the complementary energy landscape will be shown. For the purpose of visualization, a 2-dimensional Sn$_3$O$_6$ structure 
will be used as a test case.

\subsection{Energy expressions}

\begin{figure}[h]
    \includegraphics{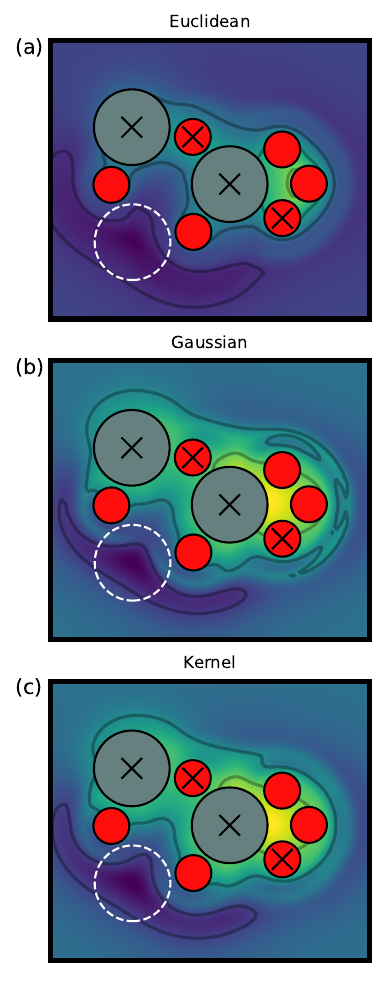}
    \caption{2-dimensional Sn$_3$O$_2$ cluster: Starting from the GM,
      one Sn is removed and the complementary energy landscape is
      inspected as a function of where the Sn atom is reintroduced.
      Three different CE expressions are considered: Using
      \textbf{(a)} Euclidean distances, $CE_{k=1}$, \textbf{(b)} Gaussian
      functions, $CE_{Gauss}$, or
    \textbf{(c)} a Kernel type method, $CE_{\rm Kernel}$. The feature vectors of the crossed atoms are picked as attractors. Darker
    blue color corresponds to lower complementary energy. Colors of atoms: Red - oxygen, grey - tin.}
    \label{fig:energy_expressions}
\end{figure}

\begin{figure}[t]
    \includegraphics{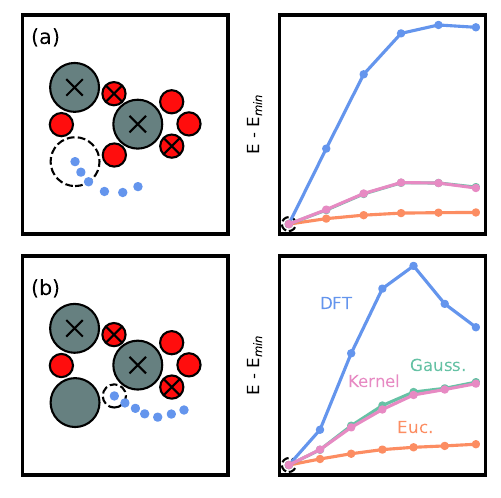}
    \caption{NEB path for moving
      \textbf{(a)} a Sn atom and
      \textbf{(b)} an O atom to the position indicated by the dashed circle, 
      and corresponding energies calculated using DFT or one of the three complementary energy expressions.
    The feature vectors of the crossed atoms are chosen as attractors for the CE energies.}
    \label{fig:energy_paths}
\end{figure}

Figure \ref{fig:energy_expressions} shows 
a Sn$_3$O$_6$ structure, where a Sn atom is missing in the bottom-left corner of the structure, indicated by a white dashed circle. The complementary energy landscape
for placing the missing tin atom will be described for different energy expressions.

In this section, inspired by previous work \cite{Rolemodel2019}, the feature vector of the $i$'th atom is calculated as

\begin{align}
    \textbf{f}_i = [\rho_i^O(\lambda), \rho_i^{Sn}(\lambda), Z_i],
    \label{eq:feature_1lamb}
\end{align}
where $Z_i$ is the atomic number of atom $i$, and $\rho_i^O(\lambda)$ and $\rho_i^{Sn}(\lambda)$ are the local densities 
of oxygen and tin atoms around atom $i$. These densitites are calculated as

\begin{align}
	\rho_i^Z(\lambda) = \sum_{j \neq i, Z_j=Z} \frac{1}{\lambda} \text{e}^{- r_{ij} / \lambda} f_c(r_{ij}),
    \label{eq:densities}
\end{align}
where $r_{ij}$ is the Euclidean distance between atoms $i$ and $j$, and

\begin{align}
	f_c(r) = 
	\begin{cases}
		\frac{1}{2} \text{cos} (\pi \frac{r}{r_c}) + \frac{1}{2},&r \leq r_c \\
		0, & r > r_c
	\end{cases}
\label{eq:cutoff_function}
\end{align}
is a cutoff function. Here, the length scale is $\lambda=1$ Å, and the cutoff distance is $r_c=10$ Å. The purpose of the large cutoff distance
is to ensure
that isolated atoms feel a force acting on them, ensuring that these isolated atoms are affected by the local optimization in the CE landscape.

The CE energy can be a function based on
the distances between feature vectors and attractors,
\begin{align}
	CE_k=\sum_{i \in I}  \min_{j \in J} ||\textbf{f}_i - \textbf{A}_j||^k ,
	\label{eq:ce}
\end{align}
where $k$ is an exponent, $I$ is the set of all atoms in the structure or a subset hereof, and $J$ is the set of attractors. Note that $k=1$ is the Euclidean distance, which was the expression used in previous work\cite{Rolemodel2019}. In this section, the set of attractors $J$
are chosen as the feature vectors of the atoms marked by crosses in Fig.\ \ref*{fig:energy_expressions}.  The $min$
operator implies that the distance is taken between a feature vector and its nearest attractor 
in this feature space. Thus, minimizing this function corresponds to lowering the distance
between a feature vector and its attractor. Since the attractors are picked as feature vectors from the present 
structure, this implies that a minimization of the complementary energy reduces the number of distinct local environments in the structure. In Nature,
low energy isomorphs often have recurring atomic motifs (think: 
high-symmetry clusters and translationally symmetric crystals), thus a generation procedure that favours structures with few distinct
local environments is a useful tactic in a global optimization setting.

The resulting complementary energy landscape using $k=1$, created by
varying the position of one tin atom, calculating feature vectors of
all atoms, picking the attractor for each atom, and evaluating the CE energy, can
be seen in Fig.\ \ref*{fig:energy_expressions}(a). The atrractors are picked as the feature vectors of the crossed atoms. These are calculated for the structure
where the Sn atom sits in the position marked by the dashed circle, and they are kept fixed for all energy evaluations.
In this figure, a dark blue color corresponds to a low complementary energy, and yellow corresponds to high complementary energy.
The global minimum of the energy landscape is found at the bottom-left part of the figure, where the Sn atom
will be bound to two oxygen atoms. 

Another possibility is to use Gaussian-type functions,
\begin{align}
    CE_{\rm Gauss}=-\sum_{i \in I} \max_{j \in J} \left[\text{exp}\left(\frac{-(\textbf{f}_i- \textbf{A}_j)^2}{2 \sigma^2} \right) \right],
    \label{eq:ce_gaussian}
\end{align}
where $\sigma=0.1$ Å is chosen. The Gaussians provide a similarity
measure between the environments and the attractors, and the
\textit{max}-operator selects the highest similarity. The leading
minus sign converts the expression to a potential energy-type expression, where the
lowest energy coincides  with the largest degree of similarities for
all atoms.
Figure
\ref*{fig:energy_expressions}(b) shows the resulting complementary energy landscape.
Compared to Fig.\ \ref{fig:energy_expressions}(a), the CE landscape includes more local minima,
but the global minimum is still found at the same position. By introducing new local minima to the CE landscape,
a local minimization of a structure in the landscape could potentially change the quality of the structure, compared to 
optimizing in a simpler CE landscape. Later in this paper,
different energy expressions will be compared to investigate the effect of changing the CE landscape.

Instead of taking only a single similarity contribution from one attractor per feature vector, it is also possible to make a kernel-like expression
in which all
the available attractors contribute, i.e.
\begin{align}
    CE_{\rm Kernel}=-\sum_{i \in I} \sum_{j \in J} \text{exp} \left( \frac{-(\textbf{f}_i- \textbf{A}_j)^2}{2 \sigma^2} \right),
    \label{eq:ce_kernel}
\end{align}
with $\sigma=0.1$ Å. The resulting $CE$ landscape can be seen in Fig.\ \ref*{fig:energy_expressions}(c). Compared to the $CE$ landscapes in Fig.\ \ref*{fig:energy_expressions}(a) and (b), this energy landscape also produces
the same global minimum. The CE landscape resembles the one in Fig.\ \ref*{fig:energy_expressions}(b), but some changes can be seen.

To see the usefulness of the simple CE expressions, two Nudged Elastic Band\cite{neb_2000} (NEB) paths have been calculated using DFT. The DFT settings are the same as those used in sec. IV.B.1. The paths and corresponding energies can be seen in Fig. \ref{fig:energy_paths}.
For the Sn atom, a small energy barrier is found, while a larger barrier is present for moving the O atom. Using the energy expressions introduced in this section,
the corresponding energies in the CE landscapes have been calculated. In both cases, the barrier is removed using eq. \eqref{eq:ce}, while eq. \eqref{eq:ce_gaussian} and \eqref{eq:ce_kernel} removed
the barrier for moving the O atom. This shows that barriers, that are present in DFT, can be avoided using CE landscapes, thereby making it possible
to obtain low energy structures via local structure optimization in the CE landscapes.

Using the simple expressions above gave different energy landscapes with the same global minimum, indicating that the
expressions can produce useful complementary energy landscapes in the context of global optimization. This highlights that it is possible to create a variety
of reasonable complementary energy landscapes without needing a complex expression for the complementary energy.

\subsection{Descriptors}
\label{sec:descriptors}

\begin{figure}[t]
    \includegraphics{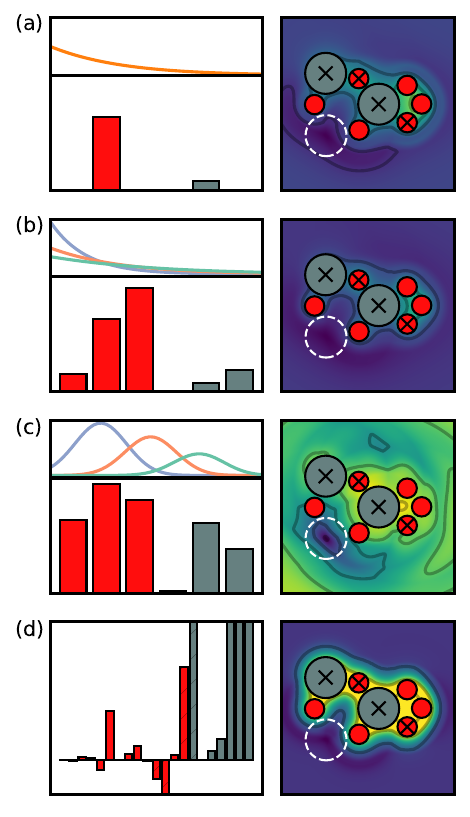}
    \caption{Different choices of descriptors and resulting CE$_{k=1}$
      landscapes. The histograms depict the feature vectors for a Sn
      atom at the white dashed circle for descriptors involving:
      \textbf{(a)} an exponential with one $\lambda$ value, $f_i$,
      \textbf{(b)} exponentials with three $\lambda$ values, $f_i^3$,
      \textbf{(c)} Gaussians, $f_i^{\rm RSF}$, and \textbf{d)} SOAP.
    The feature vectors of the crossed atoms are chosen as attractors. Note that the landscape in \textbf{a)} is the same as in Fig.\ \ref{fig:energy_expressions}a)}
    \label{fig:descriptors}
\end{figure}

The choice of descriptor influences the $CE$ energy landscape, and different descriptors can introduce
different local minima. Consider the expression for the feature vector given in eq. \eqref{eq:feature_1lamb}, where only
one value is used to describe the local density of each atomic species. This expression can easily be extended
to include more density values, which can be controlled by the hyperparameter $\lambda$. For example, using three values $\lambda_1$, $\lambda_2$, and $\lambda_3$ of 0.5 Å, 1.0 Å, and 1.5 Å, respectively,
the feature vector can be evaluated as
\begin{align}
    \textbf{f}_i^3 = [\rho_i^O(\lambda_1), \rho_i^O(\lambda_2), \rho_i^O(\lambda_3), \rho_i^{Sn}(\lambda_1), \rho_i^{Sn}(\lambda_2),  \rho_i^{Sn}(\lambda_3), Z_i],
    \label{eq:feature_3lamb}
\end{align}
where the superscript 3 indicates that three values of $\lambda$ have been used to construct the feature vector.
A reason for extending the feature vector, thereby making it a more complex descriptor, is to increase the chance
of having a unique description for different local environments, which is a desired property of the descriptor.\cite{Uniqueness2016}
If the descriptor is not unique, different local environments can be described by the same feature vector, and unwanted local minima might show up in the complementary
energy landscape.

In Fig.\ \ref*{fig:descriptors}(a) and (b), the left column shows the functions used to evaluate the feature vectors from eq. \eqref{eq:feature_1lamb} and \eqref{eq:feature_3lamb}. The histograms 
show the entries in the feature vector of the Sn atom when it is placed at the location indicated by the white dashed circle.
The last entry, $Z_{Sn}=50$, is omitted for clarity. The red and grey colors correspond to oxygen and tin densities, respectively.
The right column shows the $CE$ landscape when placing the remaining Sn atom, calculated using eq. \eqref{eq:ce} with $k=1$, where the feature vectors of the crossed atoms
are used as attractors. There is not a large difference between the two $CE$ landscapes, but for other structures the energy landscapes could potentially differ a lot.

There also exist other local descriptors in the literature that can be used when constructing complementary
energies. One example is the radial symmetry functions (RSF) \cite{Behler2011},
\begin{align}
    G_i^Z(R_s) = \sum_{j \neq i, Z_j=Z} e^{- \eta (R_{ij}-R_s)^2} f_c(R_{ij}),
    \label{eq:bp_gauss}
\end{align}
where $\eta=0.5$ Å$^{-2}$ is a hyperparameter, $R_s$ is the center of the Gaussian, and $R_{ij}$ is the distance between
atoms $i$ and $j$. Using RSFs with different values of $R_{s_1}$, $R_{s_2}$, and $R_{s_3}$ equal to 1 Å, 3 Å, and 5 Å, respectively,
a feature vector can be constructed as,

\begin{align}
    \nonumber \textbf{f}^{\rm RSF}_i = [&G_i^O(R_{s_1}), G_i^O(R_{s_2}), G_i^O(R_{s_3}), \\ &G_i^{Sn}(R_{s_1}), G_i^{Sn}(R_{s_2}), G_i^{Sn}(R_{s_3}), Z_i].
    \label{eq:feature_rsf}
\end{align}

Compared to decaying exponential functions,
which do not obtain much contribution from atoms far away from the considered atom, the Gaussians can get more information from these far-away atoms since they can be centered at different distances
away from the considered atom. Fig.\ \ref*{fig:descriptors}(c) shows the Gaussian functions, the feature vector of the Sn atom, and the complementary energy landscape
when using eq. \eqref{eq:feature_rsf} to construct the feature vectors. The energy landscape is much more complex compared to the ones seen in Fig.\ \ref*{fig:descriptors}(a) and (b),
but the GM is still found at the lower left part of the structure.

It is also possible to use more complex local descriptors, for example the Smooth Overlap of Atomic Positions \cite{SOAP2013} (SOAP) descriptor, implemented in Dscribe\cite{himanen_dscribe_2020}, which can include both
radial and angular terms. An example is given in Fig.\ \ref*{fig:descriptors}(d), where the SOAP descriptor is calculated using 
the SOAP arguments $n_{max}=3$ and $l_{max}=0$. Only radial terms have been used for better comparison with the above descriptors. Again, the same position of the GM is obtained using this descriptor.

This section has introduced some choices of different ways to calculate the feature vectors. The choice of local descriptor has an 
influence on the resulting complementary energy landscape, but seeing that different descriptors of varying complexity give the same global minimum
indicates that even simple descriptors can construct reasonable $CE$ landscapes. The choices of descriptors are of course not limited to those described here,
but choosing simple descriptors is generally desired, since 
the aim is to construct computationally cheap landscapes that can generate plausible structural candidates for the GO algorithm.

\subsection{Attractors}\label{sec:attractors}

\begin{figure}[t]
    \includegraphics{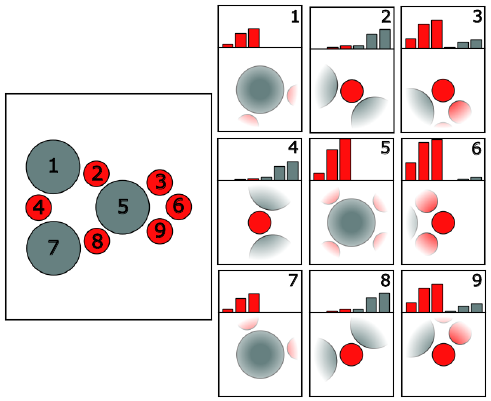}
    \caption{A Sn$_3$O$_6$ structure and its local environments, calculated using eq. \eqref{eq:feature_3lamb}.
    Because of symmetry, some of the local environments are identical.}
    \label{fig:local_environments}
\end{figure}

\begin{figure}[t]
    \includegraphics{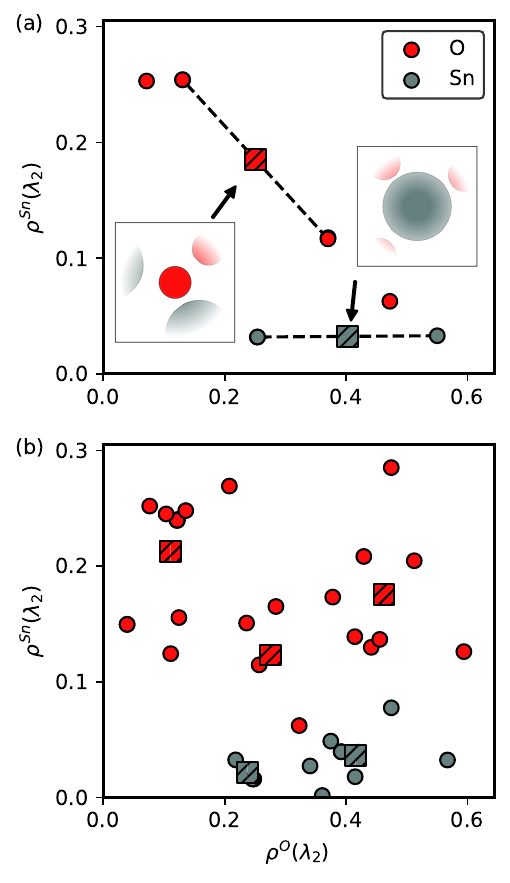}
    \caption{\textbf{(a)} The scatter plot shows the oxygen density and tin density where $\lambda_2=1.0$ Å for the nine
    local environments in Fig.\ \ref*{fig:local_environments}. Note that some of the local environments are identical.
    The squares represents attractors constructed using interpolation,
    the insets showing the possible local
    environments in cartesian coordinates for such attractors. \textbf{(b)} The oxygen density and tin density for the 
    local environments of four different Sn$_3$O$_6$ structures. 
    The squares represents attractors constructed using a k-means algorithm on all environments.
    Colors: Red - oxygen, grey - tin.}
    \label{fig:features_kmeans}
\end{figure}

Ideally, if the solution to the global optimization problem
is known beforehand, then the feature vectors from the GM should be the best choice of attractors,
since these feature vectors correspond exactly to the local environments present in the GM structure. For unsolved GO problems, where the GM structure
is not known beforehand, the choice of attractors must be related to information that is presently at hand.
In the previous sections, the set $J$ of attractors was chosen as a subset of the feature vectors in the present structure, but
other choices are also valid. In this section, a few ways to determine the set of attractors are described.

Consider the structure Sn$_3$O$_6$ in Fig.\ \ref*{fig:local_environments}. Here, all the local
environments are calculated using eq. \eqref{eq:feature_3lamb} and the feature vectors are plotted as histograms. The last entry consisting of
the atomic number has been omitted for clarity. In Fig.\
\ref*{fig:features_kmeans}(a), a scatter plot shows the second and
fifth feature in $f_i^3$ for each atom Fig.\ \ref*{fig:local_environments}.
These two features correspond to the oxygen
and tin densities calculated using $\lambda_2=1.0$ Å. Another such plot can be seen in Fig.
\ref*{fig:features_kmeans}(b), where the local environments from four different Sn$_3$O$_6$ structures are plotted. In both plots,
the color of the points represents the last entry of the feature
vector, i.e.\ the chemical identity of the atom. Many different ways to pick attractors can be envisaged. We
will formulate four ways, that can all be illustrated using Fig.\
\ref*{fig:features_kmeans}. They are:

\begin{enumerate}
    \itemsep0em
    \item \textbf{Pick from current structure:} A random number of feature vectors already present in the considered structure are chosen as attractors, 
    corresponding to picking some of the points from Fig.\ \ref*{fig:features_kmeans}(a). When minimizing the $CE$ expression, this corresponds
    to reducing the number of local environments already present in the structure.
    \item \textbf{Pick from another structure:} Pick a random number of feature vectors from another structure as attractors. This resembles a 
    mutation in the context of genetic algorithms, where information from different structures are combined. This corresponds to picking points from
    Fig.\ \ref*{fig:features_kmeans}(b) as attractors.
    \item \textbf{Interpolation:} An attractor $\textbf{A}$ is calculated by taking two feature vectors from the current structure, $\textbf{f}_i$ and $\textbf{f}_j$, and letting 
    $\textbf{A}=\frac{1}{2}(\textbf{f}_i+\textbf{f}_j)$. Two such attractors are shown as squares in Fig.\ \ref*{fig:features_kmeans}(a). The insets
    shows local environments that could correspond to such feature vectors. In this case, the atoms are guided towards local environment that are not present in the structure.
    \item \textbf{K-means:} Local environments from $N_S$ structures are collected, and a k-means algorithm is
    used to determine $N_A$ clusters, whose cluster centres will be used as attractors \cite{Clustering2018}. This method can be seen as a more complicated interpolation method
    that takes more local environments into account when creating attractors. The method is illustrated in Fig.\ \ref*{fig:features_kmeans}(b) with $N_S=4$ and $N_A=5$.
\end{enumerate}

It is important to note that after the set $J$ of attractors is determined, it remains fixed throughout the
minimization of the CE expression. That is, even though all local environments, and thus all feature vectors, change during the relaxation,
the attractors themselves do not change after they have been
constructed. We will be choosing a random number of attractors in the
following, since that adds variation in the constructed CE landscapes, which
eventually leads to the generation of more diverse and hence more
useful candidates for the search algorithm. In the following, the amount of attractors picked by 
one of the above methods will be a random number between 3 and 6, unless stated otherwise.

The above described ways of constructing attractors is by no means an
exhaustive list, and it is indeed possible to come up with other strategies.
Since the choice of attractors, like the choice of energy expression and descriptors, also influences the $CE$ landscape, there exists a lot of different ways to construct a complementary
energy landscape. It is in general difficult to know beforehand which combination of energy expression, descriptors and attractor methods
will be best suited for a given global optimization problem, so it could be useful to try out different combinations.

\section{Candidate generation with CE}
\begin{figure*}[t!]
    \includegraphics{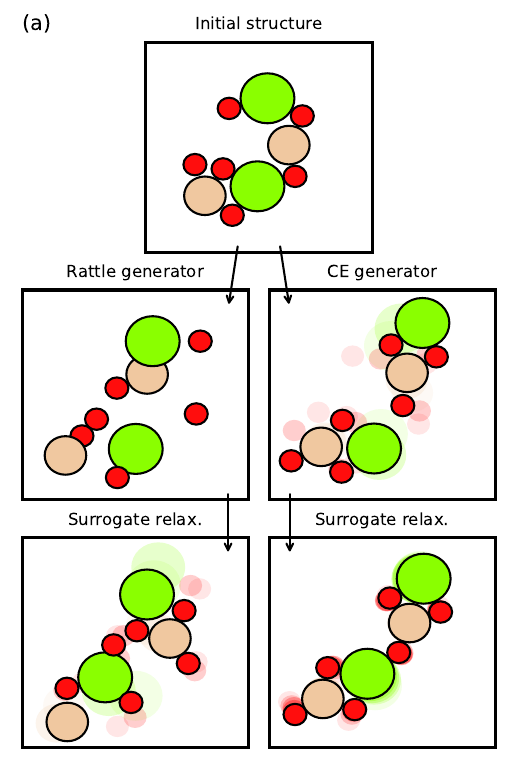}
    \hspace{0.4 cm}
    \includegraphics{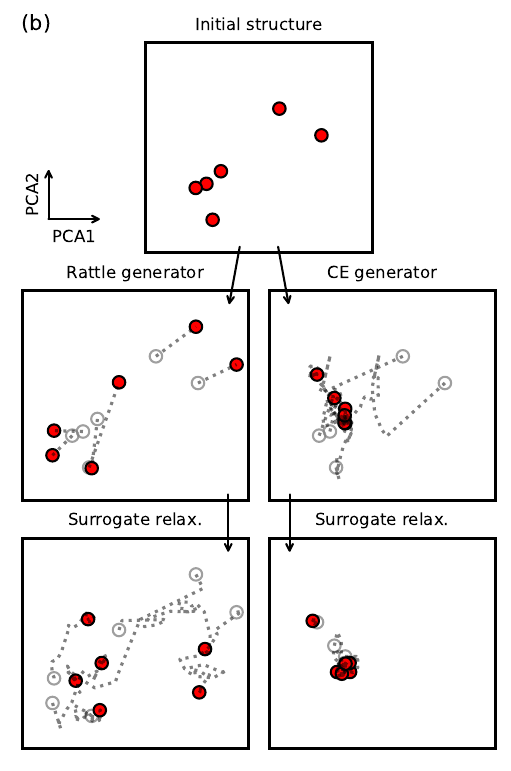}
    \caption{\textbf{(a)} The figure shows the initial (MgSiO$_3$)$_2$ 2D structure and how either a rattle generator or a CE generator, followed by a relaxation in the surrogate model, can alter the structure.  The relaxation trajectories are plotted using less intense colors.
    Colors: Red - oxygen, green - magnesium, brown - silicon.
    \textbf{(b)} The first two components of a PCA of the oxygen atoms in the different structures. Dashed lines indicate the relaxation trajectories. Grey circles represent the starting points for the relaxations.}
    \label{fig:rattle_rolemodel_structure}
    \end{figure*}

In the previous sections, we have introduced how to construct
complementary energy landscapes. In this section, we will be
illustrating how they can be used in a search context. Our discussion
will build on the Atomistic Global Optimization X
(AGOX)\cite{AGOX2022} framework and rely on its terminology. AGOX is a
python package for implementing atomistic structure search
algorithms. It supports algorithms that start from no prior
information except a method to calculate the energy and forces of a
structure in a \textit{target potential}, e.g.\ the true energy landscape, for
instance calculated with density functional theory. In AGOX, this method
is implemented as an atomic simulation environment (ASE) calculator\cite{hjorth_larsen_atomic_2017}.

AGOX is highly modularized and treats the searches as 
sequential operations on a pool of candidates kept in a temporary
storage. The operations may further access any structures kept in a permanent
storage. Such structures will often be the ones that have been
considered earlier in a search and have been treated at the target
potential level. While AGOX manipulates the pool of new candidates it may
well subject them to relaxations in other potentials, e.g.\ the CE.

A core step in a search algorithm is that of stochastically perturbing
a known structure in order to perform an explorative move. Often, such
a perturbation is carried out by randomly selecting some atoms from a structure and displacing each of
them uniformly within a disk (or a sphere for 3D problems). In AGOX,
this approach is implemented as a standard Rattle generator. With the
advent of CE landscapes, we have coded a new type of generator for
AGOX. It proceeds by:
\begin{itemize}
\item
  choosing a previously considered structure,
\item
  constructing a complementary energy landscape based on that
  structure (and possibly others, depending on the CE landscape prescription),
\item
  relaxing the chosen structure in that CE landscape.
\end{itemize}
We shall be referring to this as a \textit{CE generator}.
After generation by either the Rattle generator or the CE generator,
a search algorithm might perform a relaxation in a machine-learned potential,
which is the case for the two algorithms used in this paper. These algorithms will
be introduced in the Methods section.

Figure \ref{fig:rattle_rolemodel_structure}(a) illustrates the actions of
the Rattle generator and the complementary energy generator on a 2-dimensional silicate structure,
(MgSiO$_3$)$_2$. Starting from the initial
structure, the Rattle generator displaces some of the atoms to create
a new structural candidate for the global optimization method.  In the
new structure, a few isolated oxygen atoms are present, and some of
the other atoms are placed too close to each other, but these unwanted
features of the structure are removed by the relaxation in the
surrogate model. On the other hand, the CE generator instead produces
two MgSiO$_3$ monomers, which are combined to a single structure after
a surrogate relaxation.  In this case, the candidate generated by the
CE generator seems more reasonable in the context of global
optimization. The surrogate model used in Figure
\ref{fig:rattle_rolemodel_structure} is taken from an actual GOFEE
search and builds on 600 2-dimensional (MgSiO$_3$)$_2$ structures seen
previously in that search.

The situation can also be visualized by looking at the different steps in the feature space instead. Fig.\ \ref{fig:rattle_rolemodel_structure}(b) shows
the first two components of the principal component analysis (PCA) of the feature vectors of the oxygen atoms in the different structures from 
Fig. \ref{fig:rattle_rolemodel_structure}(a). The relaxation trajectories for the oxygen atoms are plotted as dashed lines. The PCA is based on the feature vectors
of all oxygen atoms present in the initial structure, the rattled structure, and all structures from the optimization in the complementary energy generator and the two surrogate relaxations, resulting in 822 data points in total. 
Following the Rattle generator, the oxygen atoms are spread out in feature space, which is also the case after the surrogate relaxation. 
Meanwhile, the feature vectors of the oxygen atoms are closer to each other in feature space after the CE generator has been applied to the initial
structure, and after the surrogate relaxation they are bunched together in two groups. Comparing this to the final structure in \ref{fig:rattle_rolemodel_structure}(a),
which has one oxygen atom next to a silicon atom and five oxygen atoms close to a silicon and a magnesium atom, it is clear that the CE generator, together with a surrogate relaxation,
has reduced the number of distinct local environments, which is the desired outcome of the CE generator.

\section{Methods}

\subsection{Reduced tin oxide surface}
One test case will involve finding the optimal structure of 6 Sn and 6
O atoms on a rutile SnO$_2$(110)-(4$\times$1) surface, as this has
been found experimentally to form under reducing
conditions\cite{merte_structure_2017}.

\subsubsection{Model energy}
The target potential for the tin oxide surface should ideally be a full DFT potential. However,
in order to save computational time, we have constructed a model potential that closely reproduce
the DFT and used that as the target potential. The model potential builds on a local descriptor GPR model\cite{local_model}, 
akin to the commonly used Gaussian Approximation Potential (GAP)\cite{deringer_machine_2017}.
The training data for the target model is constructed using the GOFEE algorithm. The algorithm
is described in detail later in this section. GOFEE is used to search for the GM structure of placing SnO, Sn$_2$O$_2$, and Sn$_3$O$_3$
on a SnO$_2$(110)-(2$\times$1) surface. The energy of the structures are calculated using DFT with
the Perdew-Burke-Ernzerhof (PBE) \cite{PBE1996} exchange-correlational functional and a LCAO\cite{lcao2009} basis with
a (6, 6, 1) Monkhorst-Pack k-point grid as implemented in GPAW \cite{GPAW2010}. A four-valence electron PAW setup is used for the Sn atoms.
For each stoichiometry, all structures with an energy of 10 eV above the lowest energy structure are discarded.
From here, 200 SnO, 300 Sn$_2$O$_2$, and 500 Sn$_3$O$_3$ structures are picked as training data.
The descriptor used to represent the local environments is the SOAP descriptor with $n_{max}=3$, $l_{max}=2$, and a cutoff radius of 5 Å.
To reduce the energy prediction time of the local model, the training data is sparsified using the CUR algorithm\cite{deringer_machine_2017} to obtain 1000 basis points.

\subsubsection{Basin-hopping}
The algorithm used here is Basin Hopping (BH) combined with a machine-learned surrogate model. The surrogate model,
which is trained on-the-fly during the search, is constructed using Gaussian Process Regression. Here,
instead of locally optimizing a new candidate in the target potential,
the candidate is rather relaxed in the surrogate model
and only allowed to take three optimization steps in the target potential.
The search algorithm proceeds as follows:

\begin{enumerate}
    \itemsep0em
    \item Generate a random structure and evaluate its energy.
    \item Update the surrogate model.
    \item Create a new structure using some candidate generator.
    \item Locally optimize the structure in the surrogate model.
    \item Evaluate its energy in the target potential and take three optimization steps. Add all four structures with corresponding energies as training data for the surrogate model.
    \item Use the Metropolis Monte Carlo criterion with a temperature of 1 eV to either accept or decline new candidate.
    \item Return to step 2.
\end{enumerate}

\subsubsection{Surrogate model}
This section describes the details behind the surrogate model. 
To train the surrogate model, it is useful to represent atomistic structures in a way that encodes relevant invariances, i.e. translational and rotational
invariance, and invariance with respect to swapping two atoms of the same type.
This is done using the fingerprint descriptor
by Valle and Oganov \cite{Fingerprint2010}, which is extended to include angular information. The radial part of the descriptor is calcualted as
\begin{align*}
    F_{AB}(r) &\propto \begin{cases}
        \sum_{i,j} \frac{1}{r_{ij}^2} \text{exp} \left(- \frac{(r-r_{ij})^2}{2 l_r^2} \right)&, \quad r<R_r\\
        0 &, \quad r \geq R_r
    \end{cases}\\
\end{align*}
while the angular part is constructed as

\begin{align*}
    F_{ABC}(\theta) &\propto \sum_{i,j,k} f_c(r_{ij}) f_c(r_{ik}) \text{exp} \left(- \frac{(\theta - \theta_{ikj})^2}{2l_\theta^2 } \right),
\end{align*}
where $A$, $B$, and $C$ are atomic types, $R_r=6$ Å is a radial cutoff distance, $l_r=0.2$ Å and $l_\theta=0.2$ rad are the width of the Gaussians. $f_c$ is a cutoff function of the form
\begin{align*}
    f_c(x) = 1 + \gamma \left(\frac{r}{R_\theta}\right)^{\gamma+1} - (\gamma+1)\left(\frac{r}{R_\theta}\right)^\gamma,
\end{align*}
with $R_\theta=4$ Å and $\gamma=2$. For each combination of atomic types, the radial and angular parts are binned into 30 bins and concatenated into a single feature vector.
Since the whole structure is represented using a single global feature vector, the corresponding surrogate model will be a global GPR model.
See Bisbo and Hammer for further details about the extended fingerprint descriptor.\cite{GOFEE2022}

The GPR model predicts the energy 
$E_{GPR}$ and uncertainty of the model energy $\sigma_{GPR}$ of a structure as
\begin{align}
    E_{GPR}(\textbf{x}_*) &= K(\textbf{x}_*, \textbf{X}) [K(\textbf{X},\textbf{X})+\sigma_n^2 \textbf{I}]^{-1}(\textbf{E}-\bm{\mu})+\mu(\textbf{x}_*),
    \label{eq:gpr_energy}
\end{align}
\begin{align}        
    \nonumber \sigma_{GPR}(\textbf{x}_*) =& K(\textbf{x}_*,\textbf{x}_*)- \\
    &K(\textbf{X},\textbf{x}_*)^T [K(\textbf{X},\textbf{X})+\sigma_n^2 \textbf{I}]^{-1} K(\textbf{X},\textbf{x}_*).
    \label{eq:gpr_uncertainty}
\end{align}
where $\textbf{x}_*$ is the fingerprint descriptor of the structure, $\textbf{X}$ and $\textbf{E}$ contain fingerprint descriptors and energies of the training data, $\bm{\mu}=\mu(\textbf{X})$ is the prior,
and $\sigma_n=10^{-2}$ is a noise parameter. $K$ is the kernel function, which
is a sum of two Gaussians,

\begin{align*}
    K(\textbf{x}_i,\textbf{x}_j)  =  
    &\theta_0(1-\beta) \text{exp}  \left( - \frac{1}{2} \left[\frac{\textbf{x}_i - \textbf{x}_j} {\lambda_1} \right]^2 \right) + \\ &\theta_0 \beta \text{exp}\left( - \frac{1}{2} \left[\frac{\textbf{x}_i - \textbf{x}_j} {\lambda_2} \right]^2 \right) .
\end{align*}
Here, $\beta=0.01$ is a weight, $\theta_0$ is the kernel amplitude, and $\lambda_1$ and $\lambda_2$ are different length scales, where $\lambda_1>\lambda_2$. $\theta_0$, $\lambda_1$, and $\lambda_2$ are optimized 
during the search by maximizing the log-likelihood of these values given the available data.

The prior function is defined as
\begin{align*}
    \mu(\textbf{x}) = \bar{E} + \frac{1}{2} \sum_{ij} \left( \frac{1 \text{Å}}{r_{ij} + 1\text{Å} - 0.7r_{CD,ij}}\right)^{12} \text{eV},
\end{align*}
where the sum runs over all atoms in structure $\textbf{x}$, $r_{ij}$ is the Euclidean distance between atom $i$ and atom$j$, $r_{CD,ij}$ is the sum of their covalent radii, and $\bar{E}$
is the mean of the training data energies.

\subsection{Silicate clusters}
The second test case will be that of the structure of olivine
(Mg$_2$SiO$_4$)$_4$ clusters in the gas phase. Nanosized
silicate clusters are of great interest in the astrochemistry community, as they are believed to be abundant in the interstellar medium\cite{li_ultrasmall_2001},
and the global minima of various small silicate clusters have previously been investigated\cite{woodley_structure_2009,mauney_formation_2018,Silicates2019}.
In this section, a new highly-symmetric candidate for the GM structure of the chosen stoichiometry
is being presented.

\subsubsection{DFT}
The target potential in this section is DFT with
the PBE exchange-correlational 
functional and a LCAO basis as implemented in GPAW. The 
olivine clusters are placed in a 20 Å x 20 Å x 20 Å simulation cell. The Brillouin zone is sampled using only the $\Gamma$-point.

\subsubsection{GOFEE}
The search algorithm to be used is the Global Optimization with First-principles Energy Expressions (GOFEE) algortihm,
which combines an evolutionary algorithm with the same type of global GPR surrogate model as used in the BH algorithm. In GOFEE, the surrogate model allows
for computationally cheap local optimization of structures, and it is therefore possible to generate several candidates in each iteration of the search algorithm,
relax them in the surrogate model, and pick the most promising candidate which should have its energy evaluated in the computationally expensive target potential.
The GOFEE algorithm runs as follows:

\begin{enumerate}
    \itemsep0em
    \item As a starting point, generate random structures and evaluate their energies in the target potential.
    \item Use all structures and their corresponding energies to update the surrogate model.
    \item Generate a sample consisting of the most stable, yet distinct, structures that have been evaluated in the target potential.
    \item Create new candidates using different candidate generators on the members of the current sample.
    \item Locally optimize the candidates in the lower confidence
      bound (LCB) landscape given by the surrogate model.
    \item Pick the candidate with lowest LCB energy according to the surrogate model.
    \item Evaluate energy and forces in the target potential. Take one local
      optimiztion step along these forces and evaluate the energy again. Add both structures and their eneriges as training data.
    \item Return to step 2.
\end{enumerate}
The LCB landscape is given by $E-\kappa\sigma$, where $E$ and $\sigma$
are the surrogate model mean expectation and uncertainty from Eq. \eqref{eq:gpr_energy} and \eqref{eq:gpr_uncertainty},
respectively, and where $\kappa$ is an empirical constant, often
chosen between 2 and 4.
The surrogate model used is the same as the global GPR-model described in detail in section IV.A.3 for the basin-hopping method.

\subsection{Success curves}
To give a quantitative measure of the performance of a global optimization algorithm, which is stochastic by construction, the so-called 
success curves will be used, which estimate the chance of finding the GM structure after a certain amount of
target potential evaluations.
In practice, the success curves are constructed by repeating the
search with identical settings, but with different initial structures,
a great number of times.
The success curve, $s(x)$, is then constructed by plotting the accumulated
percentage share of restarts that have found the solution after $x$
target evaluations.
Thus the usefulness of the CE generator can be judged by inspecting if
its introduction leads to higher-lying success curves.
See Christiansen et al. for more information
about the construction of success curves. \cite{AGOX2022}

\section{Results}\label{sec:results}
\subsection{Reduced rutile SnO$_2$(110)-(4$\times$1)}

This section investigates the effect of introducing the CE generator into the search algorithms. First, the machine learning assisted BH algorithm
will combine different energy expressions, descriptors, and attractor
methods in the search for the surface reconstruction of a reduced tin
oxide surface, specifically, SnO$_2$(110)-(4$\times$1)-Sn$_6$O$_6$.
Here, the CE generator will be defined by Eq. \eqref{eq:ce} with $k=1$ and \eqref{eq:feature_3lamb}, and attractors are picked from the current structure. The effect of changing either the energy expression, the descriptor, or
the attractor method while keeping the two choices will be presented. In these searches, the set $I$ of atoms that contributes to the
CE energy will be the 6 Sn and 6 O atoms that are placed on a rutile SnO$_2$(110)-(4$\times$1) tin oxide surface, and the set of attractors $J$ will be a subset of the feature vectors of these atoms.
The hyperparameters for the CE generators are the ones used in sec.  \ref{sec:complementary_section}. In searches where the CE generator is used, it is used to produce the new candidate in every second iteration. In the other iterations,
a Rattle generator is used to construct the new candidate. We note that all gradients of the CE expressions used in the local optimization of the new candidates are made
by the finite-difference method. 

\subsubsection{Energy expressions}
\begin{figure}[h!]
    \includegraphics{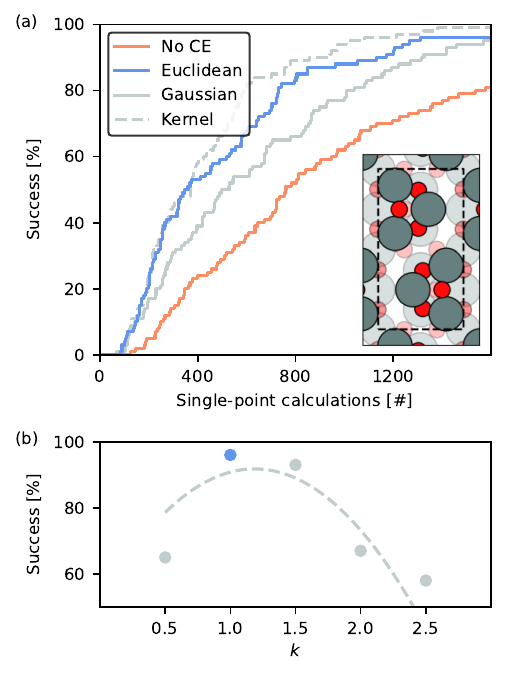}
    \caption{\textbf{(a)} Success curves using different energy expressions in the CE generator.
    In the searches behind the "No CE" orange curve, only a Rattle
    generator is used, while the CE generator replaces every second
    use of the Rattle generator in the other curves. An exponent of $k=1$ is 
    used in the blue curve. The inset shows the GM structure. The dashed line shows the supercell in which the atoms are placed.
    \textbf{(b)} The final success after 1600 target evaluations using different exponents $k$ in Eq. \eqref{eq:ce}.}
    \label{fig:success_sno2_energy_expressions}
    \end{figure}

Figure \ref{fig:success_sno2_energy_expressions}(a) shows the success curves when 
choosing different energy expressions for the CE generator. The blue curve shows the searches that use the CE generator with the equations described in the previous section. The orange curve
shows the success of searches that did not use the CE generator at all during the search for the GM. This color code will be the same for the remaining part of this paper.

The success curves show that the search can greatly benefit from
using the CE generator when constructing new candidates. This also shows that CE landscapes produced by simple energy expressions
are useful in the context of GO algortihms, meaning that one does not necessarily need complex MLPs to make structural candidates during the search for the GM structure. In \ref{fig:success_sno2_energy_expressions}(b),
the final success after 1600 target evaluations using Eq. \eqref{eq:ce} with different exponents can be seen. The choice of $k=1$, which is the blue success shown in Fig.\ \ref{fig:success_sno2_energy_expressions}(a),
gives the highest success, while other values of $k$ give lower success. We speculate that a higher value of $k$ puts more emphasis on atoms that are farthest away from their attractor,
while a smaller value of $k$ treats all atoms more evenly, and the optimal value of $k$ may therefore depend on the number of atoms and attractors for a given system.

Note that in Fig. \ref{fig:success_sno2_energy_expressions}(b), some values of $k$ give a lower final success compared to the final success of the orange curve in Fig.\ \ref{fig:success_sno2_energy_expressions}(a), which did not use the CE generator at all.
This means that based on the choices for the CE generator, the generator can decrease the performance of the algorithm, highlighting the fact that not all choices of energy expressions will improve the GO algortihm. Nonetheless, this section has shown that some choices of energy expressions can greatly improve the GO algorithm,
showing that the CE generator indeed is a good method for generation
of structures.

\subsubsection{Descriptors}
\begin{figure}[h!]
    \includegraphics{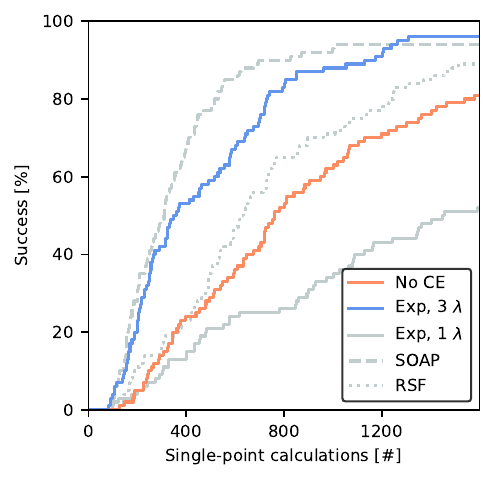}
    \caption{Success curves when using different descriptors to calculate local features.
    The "No CE" orange curve uses only a Rattle generator, while the CE
    generator replaces every second Rattle generetor in the other curves.}
    \label{fig:success_sno2_descriptors}
    \end{figure}

Figure \ref{fig:success_sno2_descriptors} shows the success curves when using the descriptors introduced in sec. \ref{sec:descriptors} in the CE generator. Here, some descriptors give rise to better success
curves than the orange one, which did not use the CE generator, while other choices of descriptors do not improve the search. Choosing only $\lambda=1$ Å, i.e. eq. \eqref{eq:feature_1lamb}, to describe the local
environment actually harms the search and gives a worse success curve compared to not using the CE generator at all. On the other hand, using 3 $\lambda$ values, RSF, or the SOAP descriptor
improves the success curves, indicating that they create useful CE landscapes when creating new candidates during the search. This also demonstrates that fairly simple descriptors
can give rise to useful CE landscapes, even though they might not be suitable for creating complex, reliable MLPs, and that there is a great variety in the choice of descriptors, even though
one has to be careful not to choose a too simple descriptor for the given problem.

\subsubsection{Attractors}
\begin{figure}[h!]
    \includegraphics{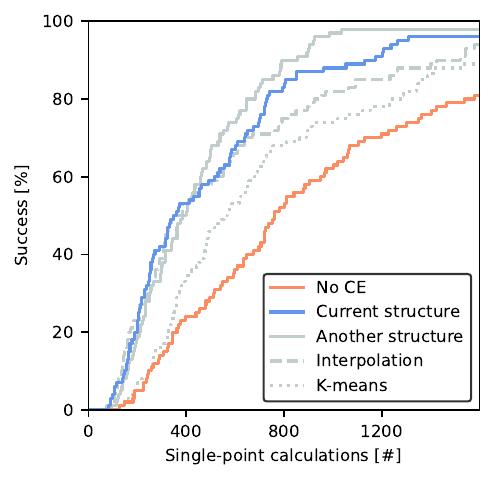}
    \caption{Success curves using different attractor mehods.
    In the "No CE", orange curve, only a Rattle generator is used. For the 'Another structure' and 'K-means' methods,
    which pick attractors from other structures, the structures are
    picked from any structure that has had
    its energy evaluated in the target potential during the search.}
    \label{fig:success_sno2_attractors}
    \end{figure}

Figure \ref{fig:success_sno2_attractors} compares success curves when using different methods to pick the attractors. It can be seen that all
choices of attractors increase the success curves, compared to the search that does not use the CE generator. Also, the methods that pick attractors
from either the current or another structure give slighty better success curves compared to the methods that creates attractors by either interpolation of k-means.
This might be explained by the fact that the attractors picked by the first two methods come from local environments that are actually present
in some structure, ensuring that they are realisable, which might not be the case for the two last methods, where completely new attractors are created. Nonetheless, this section 
also demonstrates that there exists a great flexibility in the choices regarding the CE generator, and that the GO alforithm benefits from this candidate generator.

\subsection{Olivine cluster}

\begin{figure}[h!]
    \includegraphics{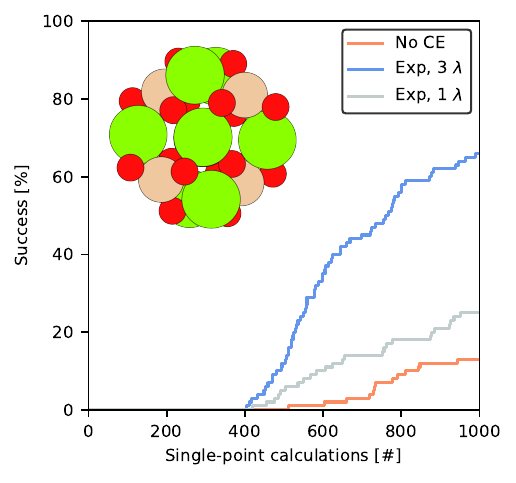}
    \caption{Success curves using the GOFEE algorithm to search for the GM of a olivine (Mg$_2$SiO$_4$)$_4$ cluster.
    In the orange curve, the CE generator is not used during the search, while the two other curves use the CE generator to generate 50\% of
    the candidates in each search iteration. The inset shows the GM structure.}
    \label{fig:success_olivine}
    \end{figure}

We end by considering a more demanding optimization problem, namely
that of finding GM of the olivine (Mg$_2$SiO$_4$)$_4$ cluster. Since
the problem is harder, we resort to the GOFEE method introduced in the
Methods section. In the present GOFEE searches, 30 new candidates are
generated in each iteration by application of random and rattle
generators. Whenever the the CE generator is introduced, 50\% of these
candidate generations are replaced with CE generator-based candidate generations.
For the CE generators, we settle on using the Euclidian distance-based
energy expression, Eq. \eqref{eq:ce} with $k=1$, and attractors that
are picked from the current structure. The set $I$ in the CE
expressions runs over all 28 atoms in the structure, and $J$ is a
subset of feature vectors for these atoms.
Two different choices for the
descriptor were made, namely the exponential descriptor using either one or three
$\lambda$ values.

Figure \ref{fig:success_olivine} presents the success curves with and
without the use of the CE generator. The low success curve without using
the CE generator testifies to the hardness of the problem.  Using the
CE generator significantly improves the chance of finding the GM
structure, especially when the slightly more complex descriptor with three
$\lambda$ values is used, compared to the simpler descriptor that only
uses one value of $\lambda$. This result further supports the statement that a more complex descriptor,
which increases the likelihood of unique description of local environments, is better suited for constrution of CE landscapes. 
Note that for this problem, using only one value of 
$\lambda$ in the exponential descriptor improved the search compared to the search that did not use the CE generator at all,
while the opposite is true for the previous problem, which can be seen in Fig. \ref{fig:success_sno2_descriptors}.
This highlights that some choice of CE generator, which increases the performance of a GO algorithm for one problem, might not be optimal for other problems.
Based on the systems considered in this work, we find that creating the CE landscape using the energy expression from eq. \eqref{eq:ce} with $k=1$,
calculating feature vectors using eq. \eqref{eq:feature_3lamb}, and picking attractors from the current structure give 
reasonable CE landscapes that help solving GO problems faster. 

The inset in Fig.\ \ref{fig:success_olivine} depicts the GM structure
found. The structure has two perpendicular 180 degree rotational symmetry axes, and it has many recurring local motifs of the atoms, which lends it
a favorable candidate for being found with a CE generator. To the best
of our knowledge, this structure has not been reported before in the
literature. Locally optimizing our new GM candidate (GM1) with the DFT
settings used in this section, such that the largest force component on an atom is 0.01 eV/Å,
the GM1 candidate is more stable by 0.45 eV than the GM candidate from ref. \onlinecite{Silicates2019} (GM2). The HOMO-LUMO gaps of GM1 and GM2 are 3.37 eV and 2.39 eV, respectively.

\section{Conclusion}
An important part of global optimization algorithms is the generation of versatile candidates that can progress the search for
the global energy minimum structure. In this paper, the general
concept of a complementary energy landscape has been discussed.
A CE generator based on the CE landscape has been implemented in the AGOX framework and used in different algorithms. Even though the complementary energy
is much simpler than MLPs, local optimization in the CE landscape can produce valuable candidates for GO algorithms, which
greatly improves the performance of different GO algorithms. We
explored the importance of different combinations of energy
expressions, local descriptors, and attractor methods used in the CE
landscape formulation, and applied a the CE generator to the problem
of the olivine (Mg$_2$SiO$_4$)$_4$ gas phase cluster, for which we
found a new global minimum energy structure.

\section{Acknowledgements}
This work has been supported by VILLUM FONDEN through Investigator grant, project no. 16562, and by the Danish National Research Foundation through the Center of Excellence “InterCat” (Grant agreement no: DNRF150).

\section{Data availability}
Version 2.3.0 of the AGOX code is available from \href{https://gitlab.com/agox/agox}{https://gitlab.com/agox/agox}. Documentation and examples 
can be found at \href{https://agox.gitlab.io/agox/}{https://agox.gitlab.io/agox/}. Training data for the energy model used in the search
for the tin oxide surface, the new GM candidate of the olivine
stoichiometry, and the data supporting our findings can be found at \href{https://gitlab.com/aslavensky/complementary_energy_paper}{https://gitlab.com/aslavensky/complementary\_energy\_paper}.

\section*{References}
\bibliography{bib}

\end{document}